\documentclass[journal]{IEEEtran}
\usepackage[dvips]{graphicx}
\begin{document}
\title{LPCH and UDLPCH: Location-aware \\Routing Techniques in WSNs}

\author{Y. Khan$^{\ddag}$, N. Javaid$^{\ddag,\$}$, M. J. Khan$^{\sharp}$, Y. Ahmad$^{\ddag}$, M. H. Zubair$^{\ddag}$, S. A. Shah$^{\ddag}$\\\vspace{0.4cm}
        $^{\ddag}$Dept of Electrical Engineering, COMSATS Institute of IT, Islamabad, Pakistan.\\
        $^{\sharp}$University of Oulu, Oulu, Finland.\\
        $^{\$}$CAST, COMSATS Institute of IT, Islamabad, Pakistan.
     }

\maketitle

\begin{abstract}
Wireless sensor nodes along with Base Station (BS) constitute a Wireless Sensor Network (WSN). Nodes comprise of tiny power battery. Nodes sense the data and send it to BS. WSNs need protocol for efficient energy consumption of the network. In direct transmission and minimum transmission energy routing protocols, energy consumption is not well distributed. However, LEACH (Low-Energy Adaptive Clustering Hierarchy) is a clustering protocol; randomly selects the Cluster Heads (CHs) in each round. However, random selection of CHs does not guarantee efficient energy consumption of the network. Therefore, we proposed new clustering techniques in routing protocols, Location-aware Permanent CH (LPCH) and User Defined Location-aware Permanent CH (UDLPCH). In both protocols, network field is physically divided in to two regions, equal number of nodes are randomly deployed in each region. In LPCH, number of CHs are selected by LEACH algorithm in first round. However in UDLPCH, equal and optimum number of CHs are selected in each region, throughout the network life time number of CHs are remain same. Simulation results show that stability period and throughput of LPCH is greater than LEACH, stability period and throughput of UDLPCH is greater than LPCH.
\end{abstract}


\IEEEpeerreviewmaketitle

\section{Introduction}

WSNs connect end users through BS or sink directly to sensor network and to provide information, according to the user need or demand ~\cite{R1}. WSN can be composed of hundreds or more sensor nodes. Which are randomly deployed inside the area of interest or very close to it and a BS or sink. Nodes sense the data and send their report toward sink. Stability period of WSN are limited. In order to prolong the stability period of WSN many routing protocols like ~\cite{R2}, ~\cite{R3}, ~\cite{R4}, etc are proposed and many new energy-efficient routing protocols must be designed. Classical approaches like direct transmission and minimum transmission energy do not guarantee well balance distribution of the energy load among nodes of sensors network ~\cite{R5}. In direct transmission, every node directly sends their data to BS, therefor far away nodes consume greater energy in sending data to BS and die quickly. However in minimum transmission energy, far away nodes send the data to BS through intermediate nodes, so nodes that are near to BS die quickly. A solution proposed is of adaptive clustering algorithm called LEACH. In LEACH, routing operation is divided in to rounds. In each round randomly CHs are selected, CHs then form clusters, and in each cluster there are cluster members and a CH. Each cluster member node sense the data and send to CH. CH receives the data, aggregate it and sends to BS. However in LEACH there is no optimum number of CHs in each round, also randomly selecting CHs makes different size of cluster; number of nodes in each cluster are vary. CH of large size cluster consume greater energy and vice versa. Therefore sensor nodes of network does not consume balance energy.

\textit{Our contribution}: In this paper, sink is not energy limited. We divide the network operation in to rounds. To improve the efficiency of WSN, we proposed two new routing protocols: LPCH and UDLPCH. These proposed protocols are clustering based techniques. In LPCH, we divide network field physically into two regions. Equal number of nodes are randomly deployed in each region. In first round, CHs are selected according to LEACH algorithm. From second round to last round, number of CHs remain same as in first round. This prolongs stability period and also greater throughput is obtained. UDLPCH follow the same thoughts to LPCH except, first round. In first round user define optimal number of CHs in each region. In case of UDLPCH stability period and throughput obtained is greater than LPCH.

\section{Related Work}
W.Heinzelman \textit{et al.}~\cite{R6} introduce a hierarchical clustering algorithm for sensor networks, called LEACH.

W.Heinzelman \textit{et al.}~\cite{R7} gives an extension of LEACH protocol uses centralized cluster formation algorithm for the formation of cluster. The algorithm execution start from the BS when the BS first receives message of all the information about each node regarding their function and energy level then it runs the algorithm for formation of CH and cluster. However, LEACH-C is not feasible for large networks because the nodes which are far away will have difficulty in sending their status to BS.

W.Heinzelman \textit{et. al}~\cite{R7} use the idea of the  clusters remain fixed and only rotate the CH with in the cluster this will increase the throughput and also saves the energy. However, the disadvantage is that the new nodes cannot be added to the network.

Yun Li \textit{et. al}~\cite{R8} improve CH selection procedure. It makes residual energy of node as the main metric which decides whether the nodes turn into CH or not after first round. In E-LEACH, first round have same probability to become CH, in next round the residual energy of each node is different after one round communication and taken into account for the selection of CHs. That means node have more energy will become CH rather than node with less energy.

Dissertation, Hang Zhou, Zhe Jiang and Mo Xiaoyan~\cite{R9} introduce a multi-hop routing protocol for WSNs called M-LEACH. M-LEACH protocol select optional path between CHs and BS through other CHs and use these CHs as a relay station to transmit data over through them. First multi-hop communication is adopted among CHs then, according to the selected optional path, these CHs transmit data to the corresponding CH which is nearest to BS. Finally CH send data to BS.

Haosong Gou and Younghwan Yoo~\cite{R10} improve network coverage by dividing network area into subareas known as partition-LEACH. In each subarea the head node is elected to receive data from other nodes with in subarea and then forwarded to BS. In partition-LEACH every node send its location and residual energy information to BS during network initialization step. The sink divide the network area into subareas according to optimal number of CH expected.

Localization problem is commonly addressed by many researchers. In localization, network ﬁeld area is logically divided into sub areas~\cite{R11}-~\cite{R12} to overcome the problem. Random deployment of nodes may cause overlapping problem, i.e. two nodes deployed in the same area may sense same location and some area may not be sensed~\cite{R13}. In~\cite{R14} authors proposed energy consumption model and tried to ﬁnd energy hole in cluster based routing protocols. They identiﬁed different areas of the network in which energy is consumed more than other areas. Similarly, authors in ~\cite{R15} proposed an energy hole removing mechanism to increase the network lifetime.

Authors in ~\cite{R16} proposed EAST protocol for WSNs. In this protocol, the temperature aware link quality estimation is done via open looping feedback and closed loop feedback process is used to reduce overhead of control packets. In ~\cite{R17} Advanced LEACH routing protocol for heterogeneous WSNs is proposed. Basically, this protocol uses static clustering i.e. at first the network area is divided into sub areas, each one acting as a cluster. Then, in each static cluster CH is selected according to the criteria set by LEACH.

Authors in ~\cite{R18} conducted a comprehensive survey based on routing protocols for WSNs. In this survey, they classified routing protocols into different categories and explored some common issues related to LEACH protocol. Moreover, how extended versions of LEACH tackled the respective issues are also discussed in this survey paper.

\begin{figure}[h]
\centering
\includegraphics[height=7cm,width=8cm]{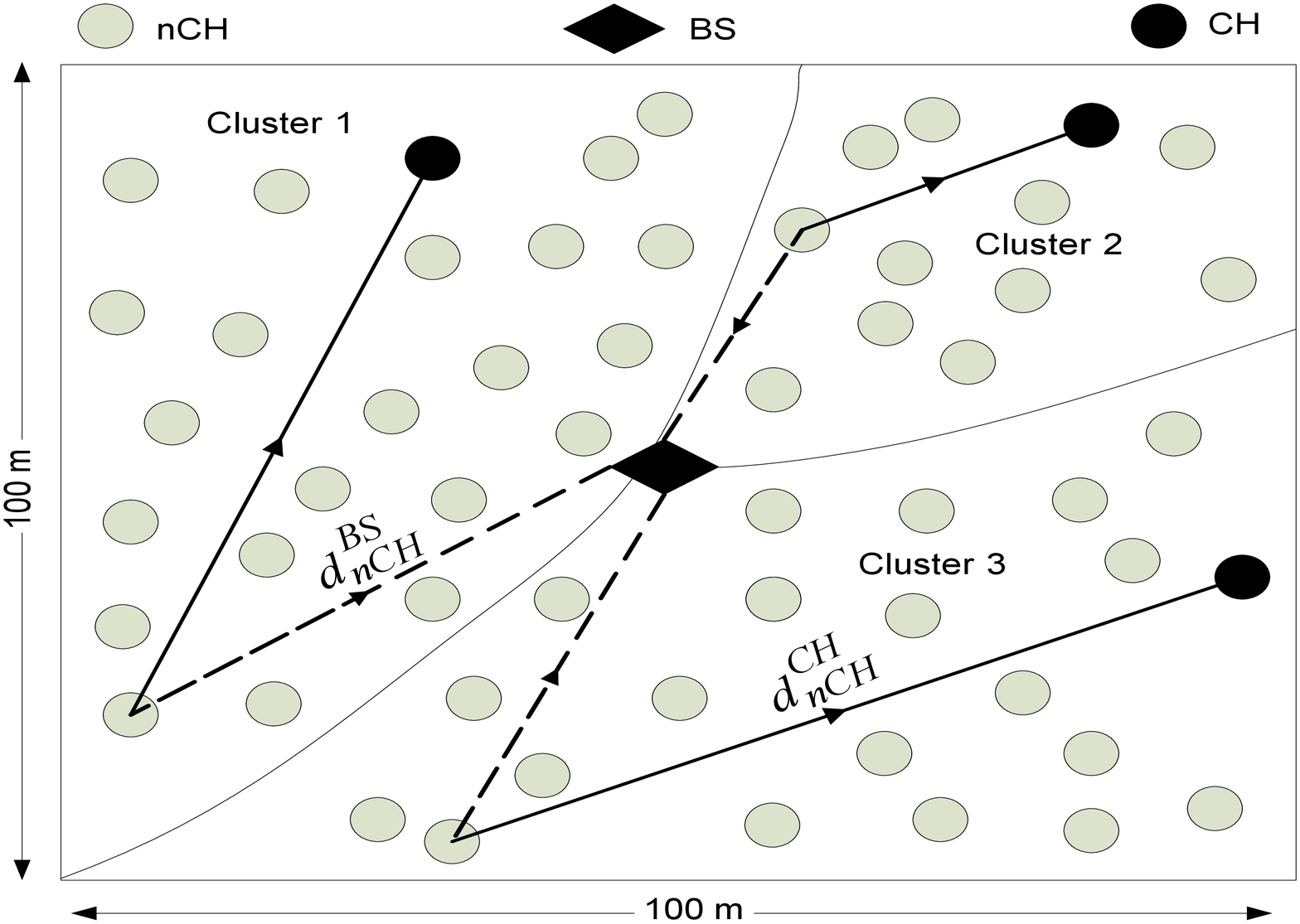}
\caption{Comparison of ($d_{nCH}^{CH}$) and ($d_{nCH}^{BS}$)}
\end{figure}
\section{Motivation}
In current research body of WSN many routing protocols are proposed. LEACH is one of the very first routing protocols. In LEACH, BS selects CHs randomly. However it can be improved in many aspects. Firstly, LEACH does not guarantee optimum number of CHs ($n\times p$) in each round of epoch, where $n$ is total number of nodes and $p$ is probability of optimal number of CHs. Secondly, CHs are selected randomly, so clusters formed are of different sizes. Therefore CH in a large size cluster (greater number of member nodes) consumes greater energy. Thirdly, member node sends data to CH, even if its distance from BS is less than its distance from CH, i.e, $d_{nCH}^{BS}<d_{nCH}^{CH}$, as can be seen in the Fig.1.

In our proposed protocols LPCH and UDLPCH, we enhance the LEACH protocol. Following the theme of~\cite{R19} and ~\cite{R20}, we divide network field physically into two regions. In LPCH, we select CHs according to LEACH algorithm where in UDLPCH, we select optimum number of CHs in each region in first round. Our proposed protocols LPCH and UDLPCH focus on controlling: cluster size, maintain number of CHs same in all rounds, distance between member nodes and CHs. which results in efficient energy consumption. Hence stability period and throughput of WSN is improved.

\section{Proposed Protocol}
In this section we discuss the network model for proposed protocols: LPCH and UDLPCH. We also discuss these protocols in detail.

\subsection{Network Model}
We divide the network field physically into two regions, and equal number of nodes are randomly deployed in each region. The BS is in the center of the field with coordinates (50, 50) shown in Fig.2. All nodes in the network are homogeneous (same energy nodes). The BS is not energy limited.
\begin{figure}[h]
\centering
\includegraphics[height=6.5cm,width=8cm]{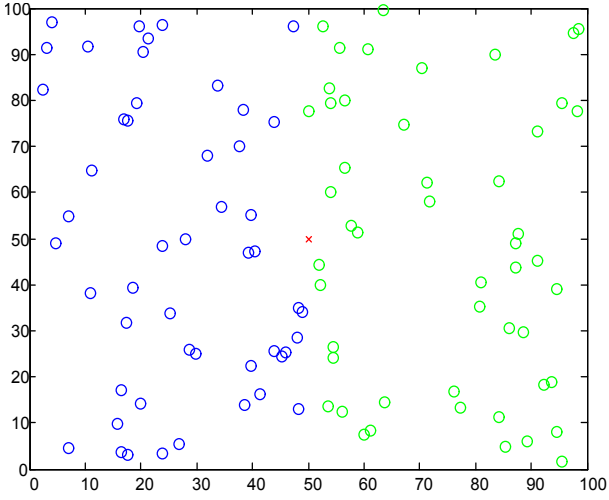}
\caption{Sensor Network Model}
\end{figure}
This model uses a location-aware clustering scheme. In location-aware scheme, BS is location aware of all nodes in the field. Each cluster has a CH which collect data from cluster members, aggregate it and send it to BS. The main features of such architecture are:
\begin{itemize}
\item
Each node has a unique ID and fixed position .
\item
Only CH node performs additional computation on the data to conserve energy of the network
\item
The sending node can adjust the transmit power to save energy depending on the distance to the receiver.
\end{itemize}
In this paper we use first order radio model, its parameters are given in table. (1). The energy consumption of the CH $(E_{CH})$ and cluster member node $(E_{NCH})$ can be calculated by following equations.

\begin{table}
\centering
\caption{First order radio model parameters}
\begin{tabular}{|l|l|}

\hline
Parameter        & Value            \\
\hline
$E_{ele}$=$E_{rec}$ & 5nJ/bit          \\
\hline        
$E_{fs}$           & 10$pJ/bit/m^2 $    \\
\hline        
$E_{amp}$          & $0.0013pJ/bit/m^4$ \\
\hline        
$E_o$            & 0.5J             \\
\hline        
Message size     & 4000             \\
\hline        
$P_{opt}$          & 0.1              \\
\hline
\end{tabular}
\end{table}

\begin{equation}
E_{rec}={lE_{ele}(n/k-1)}
\end{equation}
\begin{equation}
E_{agg}={{lE_{DA}}(n/k)}
\end{equation}
\begin{equation}
{E_{amp}^n}={lE_{amp}d_{BS}^n}
\end{equation}
\begin{equation}
E_{tra}={lE_{ele}}
\end{equation}
\begin{equation}
E_{CH}={E_{rec}+E_{tra}+E_{agg}+{E_{amp}^n}}
\end{equation}
\begin{equation}
E_{NCH}={E_{tra}+{E_{amp}^n}}
\end{equation}

Here $d_{BS}$ is the distance between CH and BS, $d_{CH}$ is the distance between CH node and member node, $E_{amp}$ is energy consumption of transceiver amplification circuit, $E_{DA}$ is data aggregation energy, $E_{ele}$ is the transmission energy of the transceiver circuit and l is the length of packet.

${E_{rec}}$ is the energy consumption by CH in receiving data from member nodes. $E_{agg}$ is the energy consumption by CH in aggregating the data. ${E_{amp}^n}$ is the net energy consumption of the transceiver amplification circuit. $E_{tra}$ is the net energy consumption in transmitting the data.

\subsection{LPCH}
In this section we explain our proposed protocol known as LPCH. In LPCH, we remove the deficiencies of LEACH protocol. We use the network model given in section 4.1. According to the model, network field is divided into two physical regions. So in our proposed protocol we change the criteria of CH selection, to select optimum number of clusters in each region. Cluster forms are of lower size. So that energy of the network is conserved. Operation of LPCH is divided into rounds. In first round, in each region CHs select according to LEACH algorithm with little modification. Because in first round all the nodes are eligible to be CH. Each node choose a random number between 0 and 1. Then, this number is compared with the threshold valve. Threshold value is set as:
\begin{equation}
T(n)=\frac{p}{1-p*(rmod(1/p))} \,\,\,\,\,\ \forall n
\end{equation}
For any node, if the number is less than or equal to threshold value, the node become CH in first round. From second round to last round, CHs select according to previous CH nodes. A node, whose y-coordinate is less than CH's y-coordinate, also its y-coordinate is closest to CH's y-coordinate select as CH for current round. When the down most node select as CH then in next round top most node become CH, this process continues throughout the network lifetime. In LPCH we improve algorithm of CH selection, for efficiently energy consumption of the network. However cluster formation is remain same as discuss in LEACH ~\cite{heinzelman2000energy}.
\subsubsection{Algorithm of LPCH}
{1}- Each node has fixed location and unique ID is assigned by algorithm given below.
\begin{figure}[h]
\centering
\includegraphics[height=3cm,width=5cm]{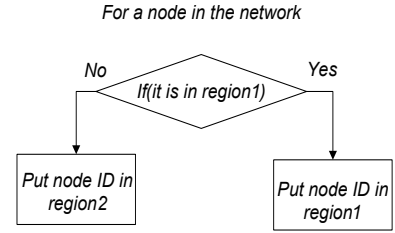}
\caption{Algorithm of putting ID to nodes in each region}
\end{figure}

{2}- In first round CHs select according to LEACH algorithm in each region.

{3}- From second round CHs select according to previous CH nodes. A node select as CH if its y-coordinate is less than y-coordinate of CH and its Y-coordinate is closest to Y-coordinate of CH as shown in Fig.4.
\begin{figure*}[t]
\centering
\includegraphics[height=12cm,width=12cm]{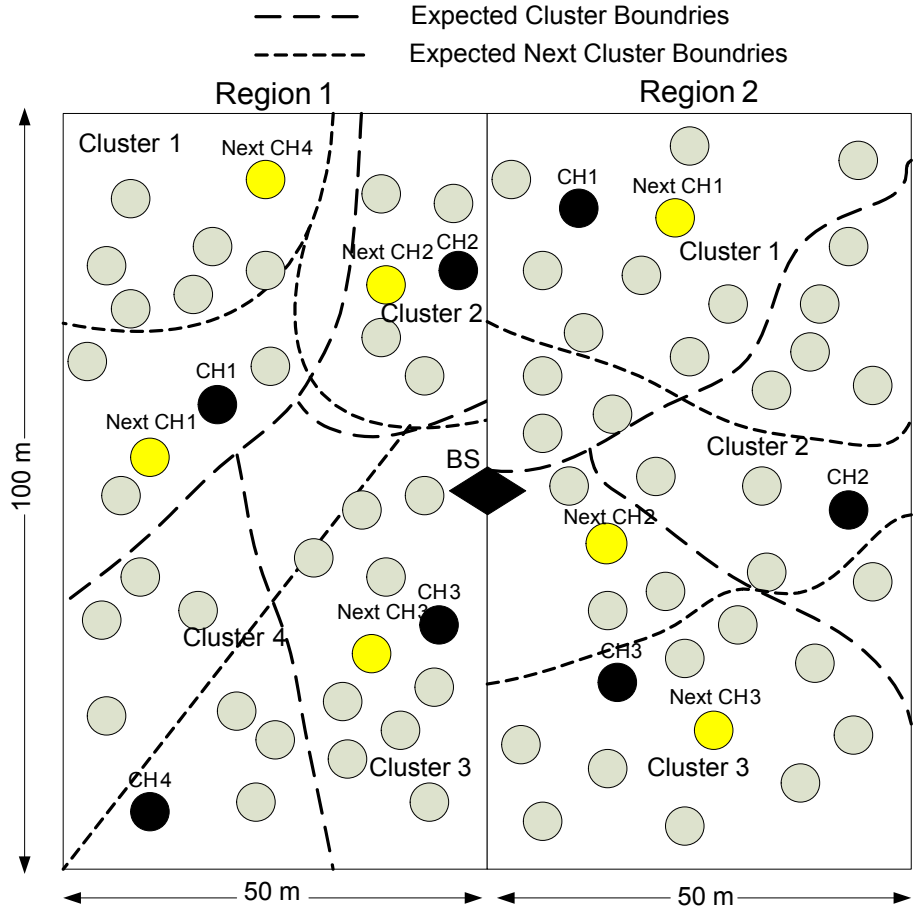}
\caption{Topology of the field for $n^{th}$ round}
\end{figure*}

{4}- If more than one node Y-coordinates are closed to CH node, then node with lowest ID become CH.

{5}- When down most node become CH in current round, then in next round the top most node become CH.

{6}- If distance of a node from BS is less than its distance from CH, it sends data directly to BS.
\subsection{UDLPCH}
In first round of LPCH; CHs are selected by LEACH algorithm in individual region, and that number of CHs remain same for all rounds. According to network model equal number of nodes are randomly deployed in each region. So in LPCH there is guarantee of selecting optimum number of CHs in the network. However, there is no guarantee of equal number of CHs in both regions. We can increase efficiency of the network by selecting equal and optimum number of CHs in each region and that number of CHs could be remain same for all rounds. We explain a new routing protocol known as UDLPCH. In first round of UDLPCH a node with ID $(i)$ become CH if $(mod(i,q)==0)$. Where $q=round(n/k)$, n is total number of nodes and k is optimal number of CHs.
For example: if $n=100$ and $k=6$ in overall network, then $q=16$. According to network model first fifty nodes are randomly deployed in first region and the remaining fifty nodes are deployed randomly in the second region. Then according to UDLPCH, nodes with ID $16,32$ and $48$ select as CHs in first region and nodes with ID $64,80$ and $96$ select as CHs in the second region. Hence, in first round equal number of CHs are selected in each region. From second round to last round, CHS select according to LPCH algorithm as discussed in the above section. Hence, throughout the network lifetime optimum number of CHs remain same in each round. 
\subsubsection{Operation of UDLPCH }
{1}- In first round equal number of CHs select in each region.
\begin{figure*}[t]
\centering
\includegraphics[height=12cm,width=12cm]{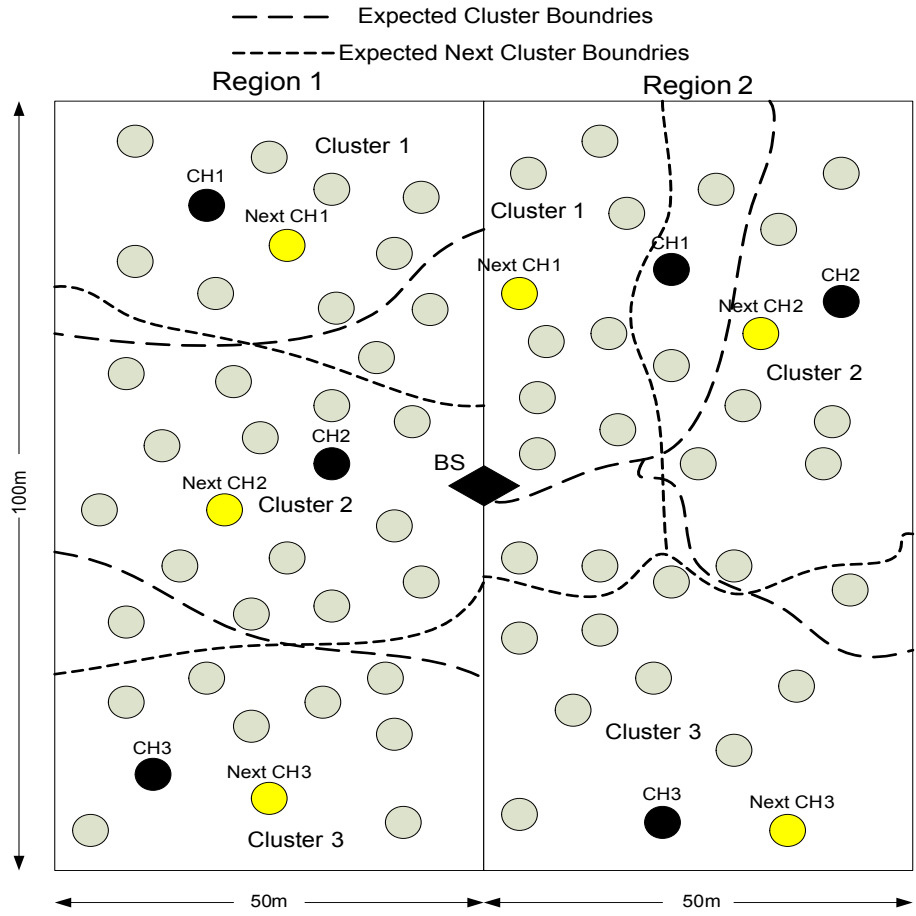}
\caption{Operation of UDLPCH}
\end{figure*}

{2}- Network coverage of UDLPCH is increased by selecting equal number of CHs in each region. Also equal size of clusters are formed.

{3}- Distance between CH and cluster member decrease, because of optimum number of CHs selection in each region. Therefore energy consumption of cluster member are reduce.

\section{Simulations and Results}
We proposed two types of clustering based protocols: LPCH and UDLPCH. For simulation of LPCH and UDLPCH we use network model given in section 4.1 with N=100 nodes and 100m x 100m field. Our goal in conducting the simulation are:
\begin{itemize}
\item
Compare the performance of LPCH, UDLPCH and LEACH on the basis of stability period and throughput .
\item
Study the effect making number of CHs constant.
\end{itemize}
Following are two subdivisions of this section. In first part LPCH is compared with LEACH and in second part we explain how UDLPCH performs better than LPCH and LEACH.

\begin{figure}[h]
\centering
\includegraphics[height=7cm,width=9cm]{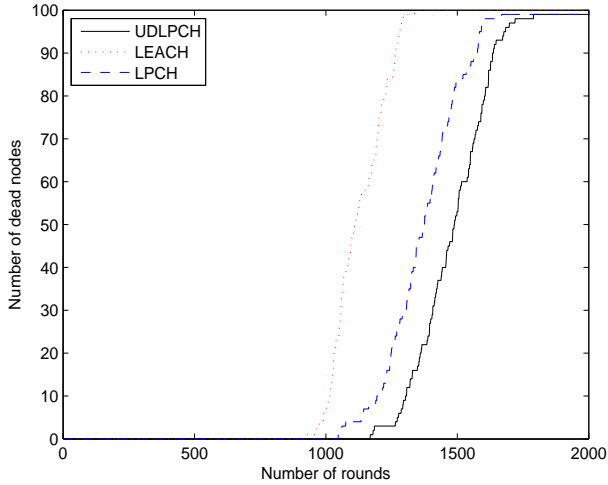}
\caption{Comparison of dead nodes in LEACH, LPCH and UDLPCH}
\end{figure}
\begin{figure}[h]
\centering
\includegraphics[height=7cm,width=9cm]{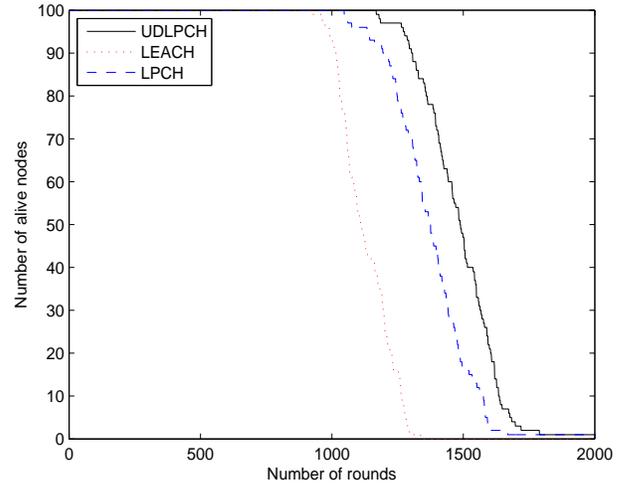}
\caption{Comparison of alive nodes in LEACH, LPCH and UDLPCH}
\end{figure}
\begin{figure}[h]
\centering
\includegraphics[height=7cm,width=9cm]{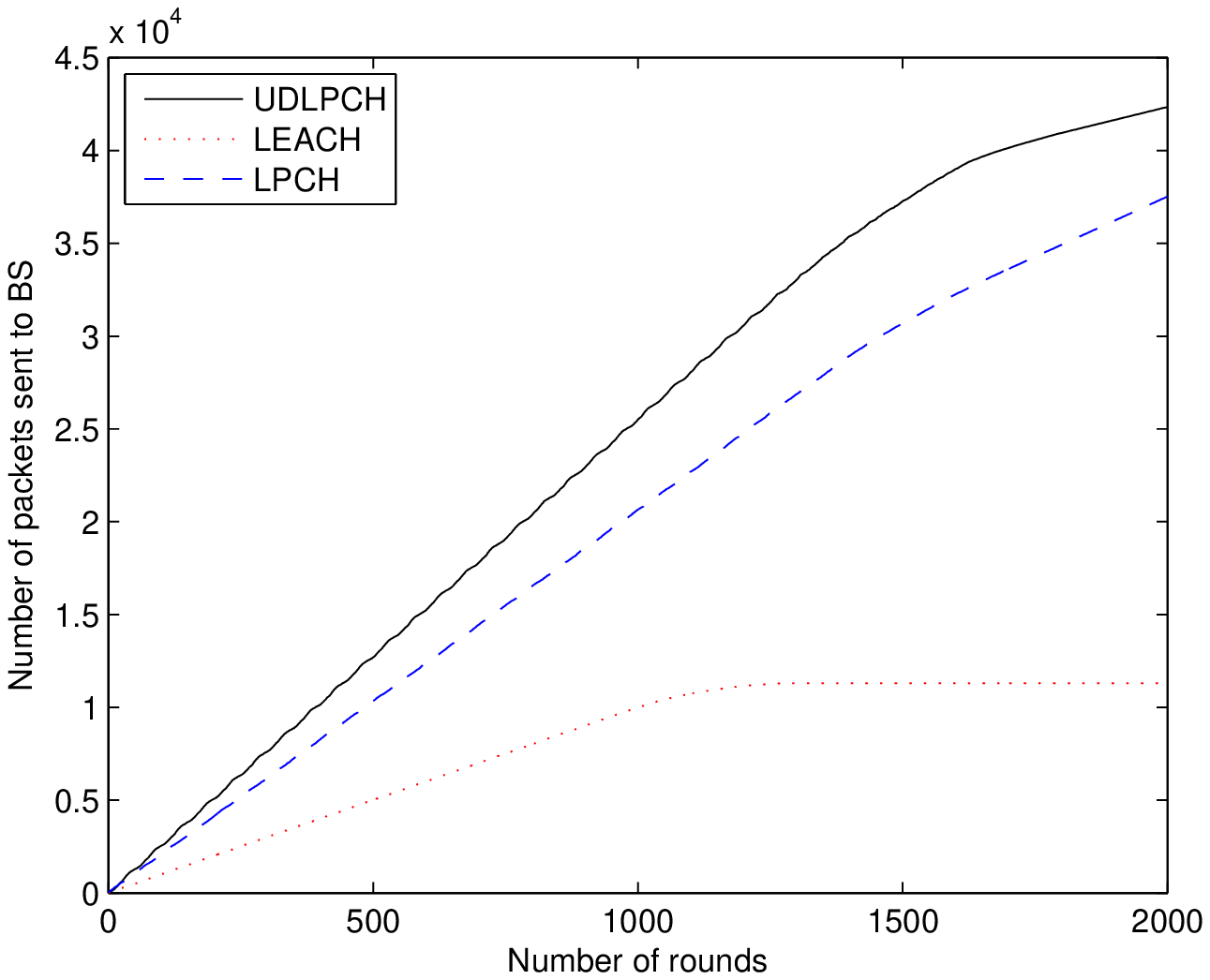}
\caption{Comparison of throughput in LEACH, LPCH and UDLPCH}
\end{figure}
{1}- We run the simulations five times, average results of simulations are shown in figures given below. Fig.6 shows number of dead nodes per round. It is obvious from the figure that, stability period of LPCH is 10\% greater than LEACH. Because, LEACH does not guarantee of selecting optimum number of CHs in each round. Where in LPCH, field is divided in to two regions and the number of CHs select in first round of each region are maintained till the end of the network.

Fig.7 shows alive nodes per round also show that network lifetime of LPCH is greater than LEACH. This is because in LEACH epoch is constant even in the unstable period. If all the nodes become CH once in first rounds of the epoch than in the remaining rounds nodes do not select as a CH in the same epoch. All nodes directly send data to BS therefore, nodes consume more energy. In LPCH there is constant number of CHs in each round. So, there are always CHs to send data of cluster members to BS therefore, energy of the network is conserved. That is why network life time of LPCH is greater than LEACH.

Fig.8 shows that total numbers of packets sent to BS in LPCH is 350\% greater than LEACH. Because in LPCH each cluster member decides to send its data to CH or directly to BS. Because BS has no power constrains. In LPCH node sends data to BS directly. If distance of a node from BS is less than its distance from CH where, in LEACH node sends data to BS if there is no CH in the running round.

{2}- As in previous section we explain that the stability period of LPCH is greater than LEACH. However, Fig. 6 shows that an average stability period of UDLPCH is 8\% greater than LPCH and 20\% greater than LEACH. Because in LPCH, CHs are selected by LEACH algorithm in first round. Hence, number of CHs select in each region may be different. The region which has more number of CHs die first. However, in UDLPCH optimum number of CHs are selected in each region, and that number of CHs remain same throughout the network.

Fig.6 and Fig.7 also show that the unstable period of UDLPCH is less than LPCH. Because in both regions equal and optimum number of CHs are selected. Therefore all nodes consumes balance energy.

Fig.8 shows that number of packets sent to BS  in UDLPCH is 12.5\% greater than  LPCH and 400\% greater than LEACH.



\end{document}